\documentclass[12pt]{article}
\usepackage{a4wide}
\usepackage{graphicx}
\usepackage{amssymb}
\usepackage{amsmath}
\usepackage{slashed}
\usepackage{cite}
\usepackage{multicol}
\usepackage{braket}
\usepackage{hyperref}
\newcommand{\be}{\begin{equation}}
\newcommand{\ee}{\end{equation}}
\newcommand{\ba}{\begin{eqnarray}}
\newcommand{\ea}{\end{eqnarray}}

\begin{document}

\allowdisplaybreaks

\begin{titlepage}
\begin{flushright}
\end{flushright}
\vfill
\begin{center}
{\Large\bf Secluded Scalar Dark Matter and the Muon Anomalous Magnetic Moment}
\vfill
{\bf Karim Ghorbani}\\[1cm]
{Physics Department, Faculty of Science, Arak University, Arak 38156-8-8349, Iran}
\end{center}
\vfill

\begin{abstract}
We consider a dark matter model with a singlet scalar, $\chi$, as our dark matter (DM) candidate which is secluded from the Standard Model (SM) and annihilates to the singlet scalar, $\phi$, via a contact interaction. 
The singlet scalar, $\phi$, has a leptophilic interaction with the SM leptons 
and may decay leptonically at tree level, and decays into a pair of photons at loop level. 
The focus in this work is to consider DM masses below 10 GeV. 
It is found a viable secluded region in the parameter space after imposing the observed 
relic density. There is a one-loop interaction between scalar dark matter and the atomic electron in this model. We then apply the available direct detection bounds from Xenon10, Xenon1T, and DarkSide on the DM-electron elastic scattering cross section. While the model can explain the muon anomalous magnetic moment, we put bounds from current and future lepton collider experiments. 
\end{abstract}

\vfill
\vfill
{\footnotesize\noindent }

\end{titlepage}

\section{Introduction}
There are two well known mechanisms for dark matter (DM) production in the early universe. One of them is called freeze-out paradigm \cite{Steigman:2012nb} where DM particles are in thermal equilibrium with the standard model (SM) particles in the early time. The second type of DM genesis is called freeze-in mechanism \cite{Hall:2009bx} in which DM particles and the SM particles have never been in thermal equilibrium.  

A prominent question about the nature of the dark matter (DM) is its interaction type with the normal matter. We may classify the DM interaction with ordinary matter into two types; interaction with atomic electrons and interaction with nucleons. 
Beside the interaction type, the mass of the DM particle is another important ingredient
in direct search for the DM particles, the so-called direct detection (DD) experiments. 
The viable range of DM mass is quite broad within the weakly interacting 
massive particles (WIMPs) paradigm \cite{Kawasaki:2013ae,PhysRevLett.64.615,Blum:2014dca}. 
In recent years direct search for dark matter with DM-nucleon interaction has been under tense scrutiny in DD underground experiments reaching unprecedented upper limits from 
LUX \cite{Akerib:2016vxi}, Xenon \cite{Aprile:2017iyp}, and SuperCDMS \cite{Agnese:2017njq}. In these experiments the most sensitivity is reached around the DM mass~$\sim 10-100$ GeV. 
In the DM mass range below 10 GeV, direct detection experiments searching 
for enticing signal from  DM-electron interaction 
become relevant, for instance, in Xenon10 \cite{Essig:2012yx}, Xeno1T \cite{Aprile:2019xxb}, and DarkSide \cite{Agnes:2018oej}.

In light of the recent confirmation of the muon anomalous magnetic moment \cite{Abi:2021gix,Muong-2:2023cdq}, new types of interactions of the SM leptons in scenarios beyond the SM are intriguing. The same type of interaction might be responsible for 
the DM-electron interaction in direct detection experiments.
The model we introduce in this work contains two singlet scalars, which are coupled 
via a contact interaction, one of which plays the role of the DM candidate and the other 
scalar has interaction with the SM leptons. This latter interaction being {\it leptophilic} is motivated from various UV complete models \cite{Batell:2016ove,Ghorbani:2021yiw,Jia:2021mwk,Chen:2015vqy,Ghorbani:2022muk}. 
The dominant annihilation of the thermal DM is into a pair of singlet scalars, 
within the so-called secluded region in the parameter space \cite{Pospelov:2007mp,Pospelov:2008jd}, 
where the DM mass is adequately larger than the singlet scalar. 
One interesting feature of the present model is that the interaction of the scalar DM with the atomic electron is induced via a one loop Feynman diagram. In a different
scenario the scalar DM which is able to account for ($g-2$) anomaly, can also have 
direct detection DM-nucleon interaction \cite{Saez:2021qta}.

The structure of the paper is the following. 
The model is described in section \ref{model}, in which the decay width of the singlet scalar is provided, and the DM-electron elastic scattering cross section, and annihilation cross section are obtained. We provide the relevant constraints in section
\ref{Constraints}. Discussions on the relic density calculation are done in section \ref{Relic-Density}, and our numerical results concerning the viable parameter space after imposing all the presented bounds are given in section \ref{Results}. We then
finish with a conclusion.

\section{The Model}
\label{model}
We consider a model which contains two scalar fields, singlet under the SM gauges, 
one acting as our DM candidate is stable and the other scalar has a contact interaction with dark matter particle. The second scalar is unstable and can decay to the SM leptons at the classical level. The relevant Lagrangian has the structure, 
\begin{equation}
{\cal L}_{\text{DM}}  = \lambda \chi^2\phi^2 + \alpha \sum_{l=e,\mu,\tau} \frac{m_l}{v}~ \phi l^+ l^- \,, 
\end{equation}
where $v = 246$ GeV is the vacuum expectation value of the SM Higgs doublet, 
and the leptophilic scalar $\phi$ interaction 
with the SM leptons has a mass-proportional strength.
This type of interaction is motivated in a UV complete model with two Higgs-doublet 
and an additional scalar singlet \cite{Batell:2016ove,Ghorbani:2021yiw,Jia:2021mwk}. 
As an alternative UV completion for this model one may consider a scenario with vector-like fermions at the weak scale \cite{Chen:2015vqy}.

The scalar $\phi$ may decay to the SM fermions via $\phi \to e^+ e^-, \mu^+ \mu^-, \tau^+ \tau^-$ with the following decay width for 
the leptonic process, $\phi \to l^+l^-$,
\begin{equation}
 \Gamma_{l^+l^-} = \frac{\alpha^2}{8\pi} \frac{m_l^2}{v^2} (1-m_l^2/m_\phi^2)^{3/2} \,.
\end{equation}
The singlet scalar with an arbitrary mass can decay to a pair of photons at loop level with the following decay width \cite{Djouadi:2005gi},
\begin{equation}
 \Gamma_{\gamma \gamma} = \frac{\alpha^2 \alpha_{em}^2}{256 \pi^3} \frac{m_\phi^3}{v^2} 
 \lvert \sum_{l= e,\mu,\tau} {\cal M}_{1/2}(s_{l}) \rvert^2 \,,
\end{equation}
where $s_l = m_{\phi}^2/4m_{l}^2$, and 
${\cal M}_{1/2} (s_l) = 2 [s_l + (s_l-1)f(s_l)]/s^2_{l}$ with

\[
f(s_l) = 
\begin{cases}
  Arc sin^2 \sqrt{s_l}  
  &  s_l \leq 1 \,, \\
-\frac{1}{4} \Big( \log \frac{\sqrt{s_l}+\sqrt{s_l-1}}{\sqrt{s_l}-\sqrt{s_l-1}} -i\pi \Big)^2
 &   s_l > 1 \,.
  \end{cases}
\]

\begin{figure}
\begin{center}
\includegraphics[width=0.25\textwidth,angle =0]{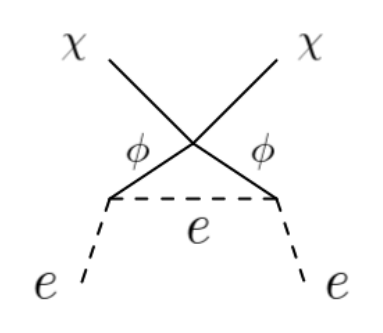}
\end{center}
\caption{Shown is the leading Feynman diagram for the elastic scattering of the DM candidate off the atomic electron.}
\label{DD-diagram}
\end{figure}

Since the scalar DM has a contact interaction with the scalar $\phi$, 
when $m_\phi \lesssim m_\chi$, then the  DM annihilation cross section times the relative velocity reads
\begin{equation}
 (\sigma_{\text{anni}} v_{\text{rel}})_{\chi \chi \to \phi \phi} = \frac{\lambda^2}{\pi s} \sqrt{1-4m_\phi^2/s} \,.
\label{anni}
 \end{equation}
This formula is needed when computing the DM relic density. The mandelstam parameter, $s$,
is the center of mass energy of the two annihilating dark matter particles.

In the present model there is no tree level interaction between the DM scalar and the SM leptons. The leading DM-electron interaction induced via a loop interaction as depicted in Fig.~\ref{DD-diagram}.
The scattering amplitude for the DM-interaction reads  
\begin{equation}
 {\cal M} = 4\lambda \frac{\alpha^2 m_e^2}{v^2} \int \frac{d^4l}{(2\pi)^4}
\frac{\bar e(p_2) (\slashed{l} + m_e) e(p_1)}{[(p_2-l)^2-m^2_\phi][(p_1-l)^2-m^2_\phi][l^2-m^2_e]} \,.
\end{equation}
Taking the form factor $F_{\text{DM}}(q) = 1$ \cite{Essig:2011nj},
the scattering cross section for DM-electron is obtained as  
\begin{equation}
 \sigma_e \sim \frac{\lambda^2 g^4_e}{128 \pi^5 m^2_{\text{DM}}} |{\cal F}(m_\phi)|^2 \,,
\end{equation}
where $g_e = \alpha m_e/v$, and for ${\cal F}(m_\phi)$ we have
\begin{equation}
 {\cal F}(m_\phi) = 1 + \frac{m^2_\phi-3m^2_e}{2m^2_e}~\text{log} (\frac{m^2_e}{m^2_\phi})
+\frac{(m^2_\phi-m^2_e)\sqrt{m^4_\phi-4 m^2_\phi m^2_e}}{m^2_e m^2_\phi}
~\text{log} \Big( \frac{m^2_\phi+\sqrt{m^4_\phi-4 m^2_\phi m^2_e}}{2m_\phi m_e} \Big) \,.
 \end{equation}
The singlet scalar has a loop contribution to the anomalous magnetic moment of the muon, the so called $g-2$ anomaly \cite{Schwinger:1948iu,PhysRevD.5.2396}. The new physics correction is obtained as  
\begin{equation}
 \Delta a_\mu = \frac{\alpha^2 m^2_\mu}{v^2} \int^{1}_{0} dy \frac{(1+y)(1-y)^2}{(1-y)^2+y (m_\phi/m_\mu)^2} \,,
\end{equation}
where the correction depends on the coupling $\alpha$ and $m_\phi$ as free parameters.
In this model there are four independent free parameters, namely, $m_\chi$, $m_\phi$, $\lambda$, and $\alpha$. We may use $m_\chi$ and $m_\text{DM}$ interchangeably.

\section{Constraints}
\label{Constraints}
In this section we collect all the constraints which are relevant for the model in this work.

\begin{itemize}
 \item {\it Relic density}: The dark matter relic abundance is measured by WMAP and Planck satellite with unprecedented accuracy, $\Omega h^2 = 0.12\pm 0.001$ 
 \cite{Planck:2018vyg}. This will put constraints on the three parameters of the model, namely, the coupling $\lambda$, and masses $m_{\text{DM}}$ and $m_\phi$.
 
 \item $(g-2)_\mu$ {\it anomaly}: There has been a long-standing discrepancy 
 on the measurement of the muon magnetic moment and its theoretical prediction within the SM \cite{Bennett:2006fi}. This discrepancy is defined as $\Delta a_\mu = a_\mu(\text{Exp}) - a_\mu(\text{SM})$. 
 A recent update provided by FNAL supports the deviation with a significant of 
 $\sim 4.2 \sigma$ \cite{Abi:2021gix}, with $\Delta a_\mu = (25.1\pm 5.9) \times 10^{-10}$. 
 The muon ($g-2$) collaboration has also released their results based on the 
 data collected in 2010 and 2020 \cite{Muong-2:2023cdq}.
 The results reads  $\Delta a_\mu = (24.9\pm 4.8) \times 10^{-10}$ at $5.1$ standard deviation, 
 indicating an improvement in precision by a factor of two. We will use an updated world average
 for this quantity in our numerical results.
 
 \item {\it BABAR experiment}: The production of a new scalar is feasible via the process $e^- e^+ \to \mu^- \mu^+ \phi$ in lepton colliders, where the scalar $\phi$ will decay into a pair of muons subsequently. The {\it BABAR} experiment have found  strong constraints on the coupling $\alpha$ in the channel with four muons in the final state \cite{TheBABAR:2016rlg}. 
  In addition, for a leptophilic scalar, $\phi_L$, which decays mainly to the SM leptons, the {\it BABAR} experiment provides constraints on the muon-scalar coupling \cite{BaBar:2020jma}. 

  \item Belle II: This is a high-luminosity B-factory which collected a significant number of $\tau^+ \tau^-$ in association with scalars. Since the scalar coupling is proportional to the lepton mass then the scalar production from taus is the dominant process, see \cite{Batell:2016ove}.  

  \item ILC as future international linear collider: This is a future lepton collider as Higgs factory which runs at the center of mass $\sqrt{s} \sim 250$ GeV \cite{ILC:2013jhg,Bambade:2019fyw}. 
  We will apply the constraints from an ILC machine at 250 GeV with integrated luminosity $2000 fb^{-1}$. 
  \cite{Chun:2021rtk}
  
  \item The Light Dark Matter eXperiment (LDMX): 
   This is a multi-purpose future experiment \cite{LDMX:2018cma} which provides a high-luminosity measurement of missing momentum. The primary goal of LDMX is to search for light dark matter in the sub-GeV mass range. We will apply the projections of LDMX in Phase.1 and Phase.2 with the measurement of the muon missing momentum (M3). 
\end{itemize}

\section{Relic Density}
\label{Relic-Density}
\begin{figure}
\begin{center}
\includegraphics[width=0.45\textwidth,angle =-90]{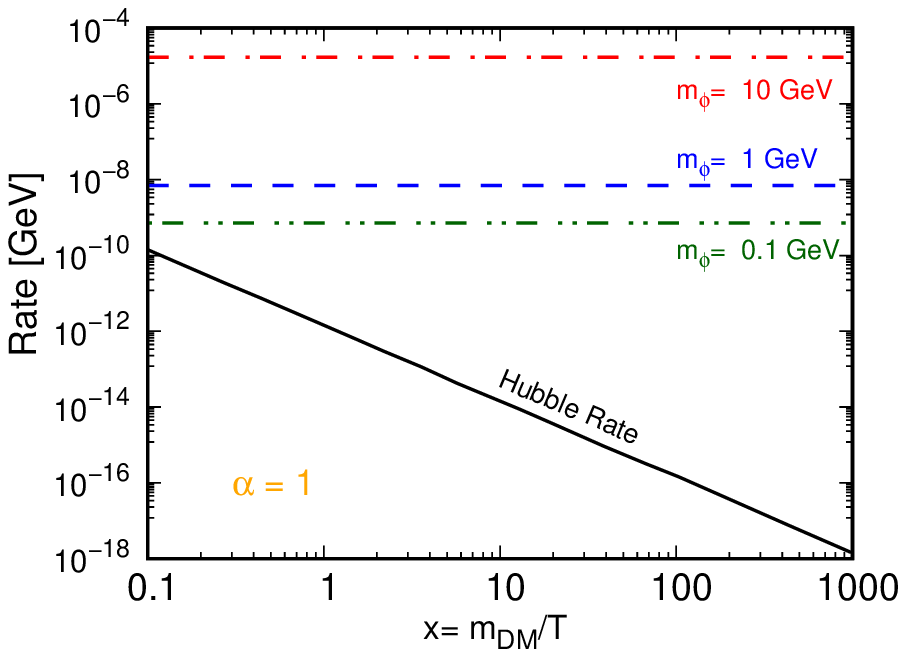}
\end{center}
\caption{The Hubble rate and the total decay width of the singlet scalar as a function of the parameter, $x$, are shown for three values of the singlet scalar mass.}
\label{Rate}
\end{figure}
In the present model there is no dominant DM annihilation to the SM particles at tree level in perturbation theory. There is rather a secluded dark matter annihilation to a pair of singlet scalars. In the thermal bath, the DM entropy is transferred via the secluded annihilation 
$\chi \chi \to \phi \phi$. The relevant interaction is a contact one with the strength proportional to $\lambda$. 
The annihilation of DM to the SM leptons is also possible via off-shell singlet scalars, however, these interactions are negligible, leaving us a secluded region in the parameter space to probe. 

The singlet scalar in the thermal bath can decay to the SM leptons when kinematically allowed. If the strength of the scalar-lepton interaction is sufficiently strong 
then it is possible to maintain the dark sector and the SM sector in kinetic equilibrium.
As long as the DM particles have strong enough interaction with the singlet scalar, 
to maintain the kinetic equilibrium, it is sufficient that the singlet scalar and the SM 
particles be in kinetic equilibrium. To this end, we constrain the total decay width 
of the scalar to be larger that the Hubble rate at the freeze-out temperature, 
$H(T_f) \leq  \Gamma (\phi \to \text{SM})$, where $T_f = m_{\text{DM}}/x_f$, and 
according to the standard lore expectation is $x_f \sim 15-20$. We examine this condition
for three scalar masses, $m_\phi = 0.1, 1$ and 10 GeV, in Fig.~\ref{Rate}, with the scalar coupling $\alpha = 1$. It is evident that in the range of interest for the scalar 
mass this condition is held. Taking larger value for $\alpha$, the condition is satisfied 
even easier. In addition, there is another condition for the scalar lifetime, 
where the scalar should decay before the BBN epoch, $t \sim 1$ sec. This can be seen from the 
scalar decay widths in Fig.~\ref{Rate}, that the BBN condition is easily fulfilled in the region of the parameter space of our interest.

Now, the secluded annihilation channel is open even when the singlet scalar mass is slightly larger than the DM mass, due to the kinematical tail in the Boltzmann distribution. 
To describe the time evolution of the number density of the scalar DM via freeze-out mechanism, one needs to solve the Boltzmann equation, 
\begin{equation}
 \frac{dn_{\chi}}{dt} +3Hn_{\chi} = 
 - \langle (\sigma_{\text{ann}}v_{\text{rel}})_{\chi \chi \to \text{SM}} \rangle [n^{2}_{\chi}-(n^{\text{eq}}_{\chi})^2 ] 
 - \langle (\sigma_{\text{ann}}v_{\text{rel}})_{\chi \chi \to \phi \phi} \rangle [n^{2}_{\chi}-(n^{\text{eq}}_{\chi})^2 (\frac{n_{\phi}}{n^{\text{eq}}_{\phi}})^2] 
 \,,
\end{equation}
where, the Hubble parameter is denoted by $H$, and the DM number density at equilibrium is $n^{\text{eq}}_{\chi}(T) = \frac{m_\chi^2}{2\pi^2}TK_2\left(\frac{m_\chi}{T}\right)$. 
Now, since in our model $(\sigma_{\text{ann}}v_{\text{rel}})_{\chi \chi \to \text{SM}}
\sim 0$, and the singlet scalar has been in thermal equilibrium with the SM plasma, then
$n_{\phi} = n^{\text{eq}}_{\phi}$. The Boltzmann equation then takes on the form,
\begin{equation}
 \frac{dn_{\chi}}{dt} +3Hn_{\chi} = 
  - \langle (\sigma_{\text{ann}}v_{\text{rel}})_{\chi \chi \to \phi \phi} \rangle [n^{2}_{\chi}-(n^{\text{eq}}_{\chi})^2 ] \,.
\end{equation}
The thermal average of the DM annihilation cross section times the velocity is given by
\begin{equation}
\langle \sigma_{\text{ann}} v_{\text{rel}} \rangle = \frac{1}{8 m_{\chi}^4TK^{2}_{2}(\frac{m_{\chi}}{T})}
\int^{\infty}_{4m^{2}_{\chi}} ds~(s-4m^{2}_{\chi})\sqrt{s}~K_{1} (\frac{\sqrt{s}}{T})~\sigma_{\text{ann}}(s)\,, 
\end{equation}
where, the modified Bessel functions of first and second rank are denoted by $K_{1,2}$.
The package Micromegas will be employed to evaluate the DM relic density numerically \cite{Barducci:2016pcb}. 
\begin{figure}
\begin{center}
\includegraphics[width=0.45\textwidth,angle =-90]{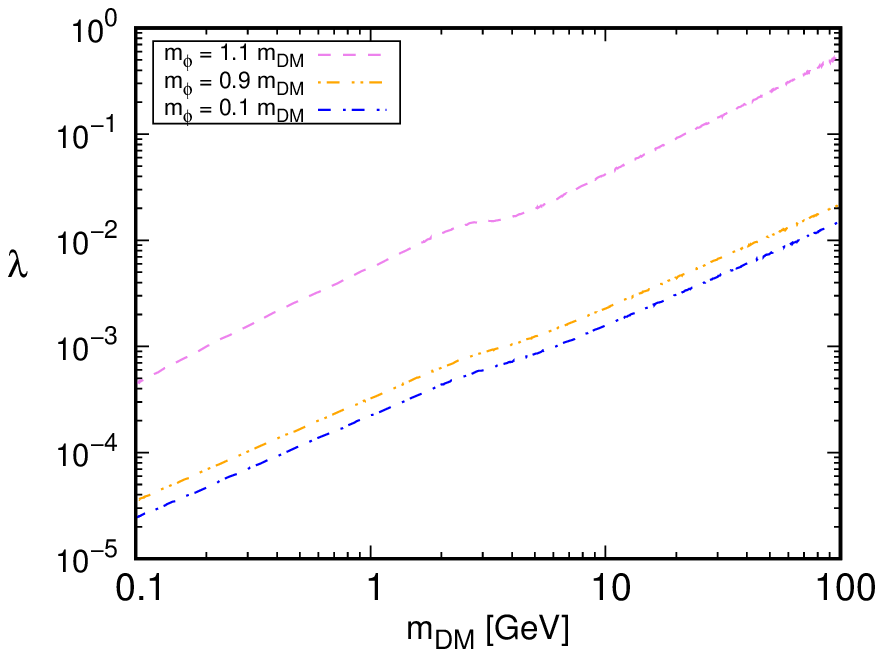}
\end{center}
\caption{The dark coupling, $\lambda$, as a function of the DM mass is given 
after imposing observed DM relic density, for three values of the singlet scalar mass.}
\label{lambda}
\end{figure}
The only coupling, $\lambda$, controls the size of the DM relic density. Thus, by imposing the observed relic density, we find the viable values for the coupling in terms of the DM mass for the singlet scalar mass $m_\phi = (0.1, 0.9,1.1) m_{\text{DM}}$. The results given 
in Fig.~\ref{lambda} show that by increasing the DM mass, larger value for the coupling is picked out. This is expected, since the annihilation cross section in 
Eq.~\ref{anni} falls off by increasing the DM mass, and thus the coupling $\lambda$ should grow in such a way to compensate the reduction of the cross section. 
The case with $m_\phi = 1.1 m_{\text{DM}}$, where the singlet scalar is 
slightly larger than the DM mass, can also give rise to the correct relic density 
since the DM velocity lies in the tail of the Boltzmann distribution.

\section{Results}
\label{Results}
\begin{figure}
\begin{center}
\includegraphics[width=0.45\textwidth,angle =-90]{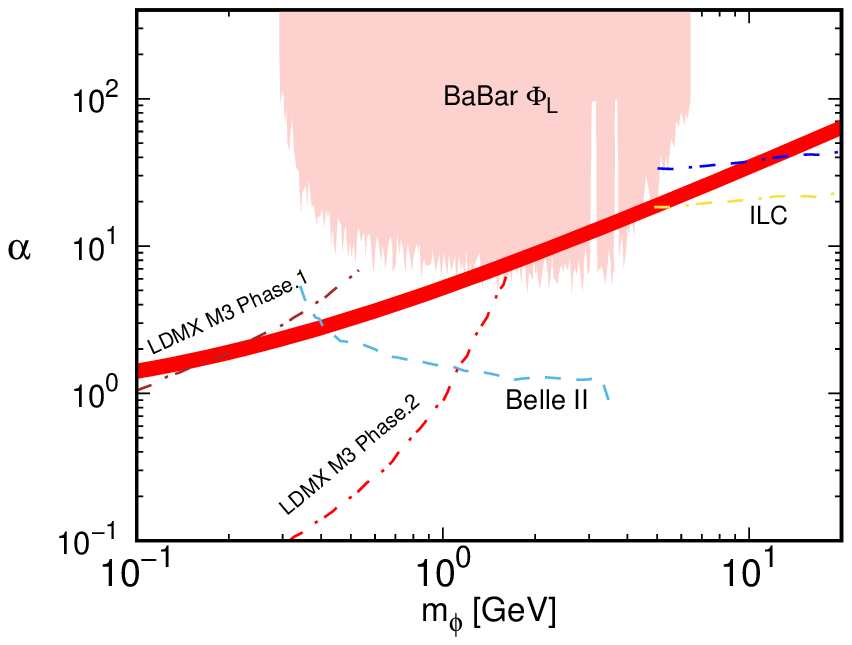}
\end{center}
\caption{The region consistent with the discrepancy in the muon ($g-2$) anomaly
is shown in the plane $\alpha-m_\phi$. The collider bounds from the current and future
experiments are placed.}
\label{exclusion}
\end{figure}
In this section we will put together all the astrophysical, collider, and direct detection bounds which are relevant for the parameter space of the present model. The aim is 
to find the viable regions which comply with discrepancy found in the muon magnetic moment. 
First we display in the $\alpha-m_\phi$ plane, the region consistent with the muon anomalous
magnetic moment, ($g-2$) band, see Fig.~\ref{exclusion}. Then we apply in the same figure a bound from {\it BABAR} experiment, which excludes the ($g-2$) band in the singlet scalar mass range 
$m_\phi \sim 1.5-5$ GeV. Along the same line Belle II, first super B-Factory experiment, is able to reach 
a significantly higher sensitivity to exclude the scalar mass range $m_\phi \sim 0.3-3.5$ GeV. As depicted in Fig.~\ref{exclusion} the expected bounds from a future experiment, LDMX, can probe regions in the parameter 
space with scalar mass up to about 0.5 GeV in its first phase and masses up to about 1.5 GeV in its second phase. 
Finally, we present constraints from ILC, as a future international linear collider, which is sensitive to scalar masses above 5 GeV. 

\begin{figure}
\hspace{-.75cm}
\begin{minipage}{0.37\textwidth}
\includegraphics[width=\textwidth,angle =-90]{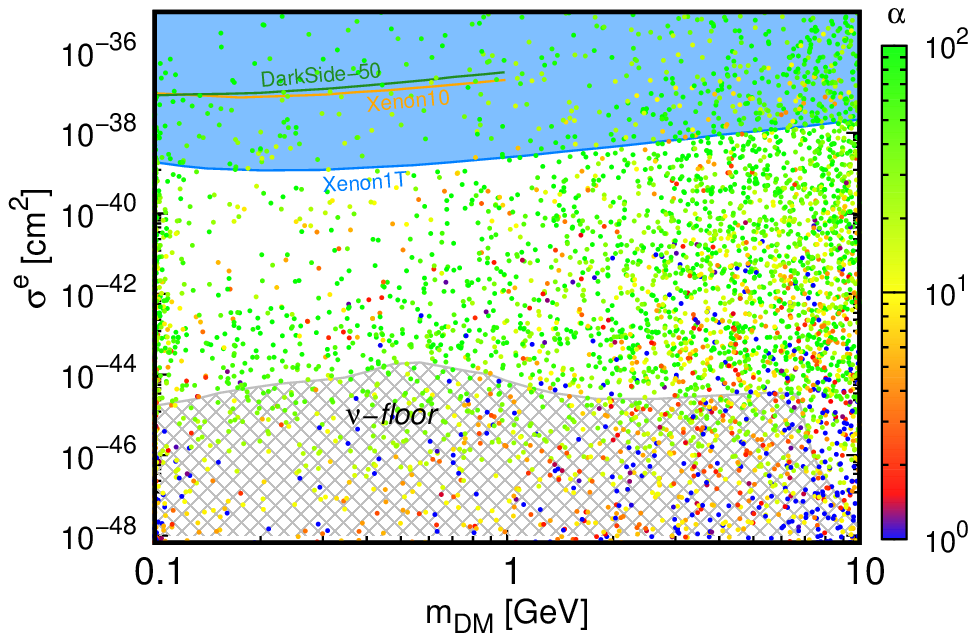}
\end{minipage}
\hspace{2.5cm}
\begin{minipage}{0.37\textwidth}
\includegraphics[width=\textwidth,angle =-90]{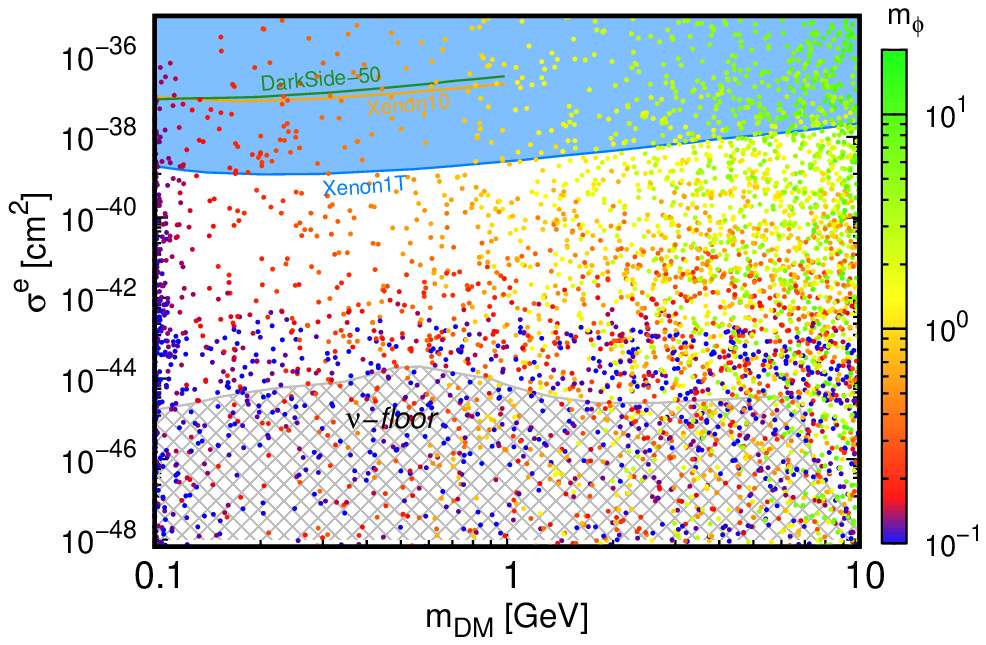}
\end{minipage}
\caption{ The DM-electron elastic scattering cross section is shown 
as a function of the DM mass. 
In the {\it left panel} the scalar coupling, $\alpha$, is indicated by 
the color spectrum vertical bar, while in the {\it right panel}, color spectrum 
shows the values for $m_\phi$. In both plots bound from the observed relic density
is imposed. Direct detection bounds from Xenon10, Xenon1T and DarkSide are placed.
As well, the neutrino floor is shown in both plots.}
\label{direct}
\end{figure}

In the next step we go on to compute the DM-electron elastic cross section as a function of the DM mass for points in the parameter space which already 
respect the bound on the relic density provided by WMAP and Planck. 
We scan regions with 0.1 GeV $< m_{\text{DM}} <$ 10 GeV, 0.1 GeV $< m_\phi <$ 10 GeV, $10^{-5}< \lambda < 1$ and $1 < \alpha < 150$. 
In Fig.~\ref{direct} we can find regions respecting the 
direct detection bounds from Xenon10, Xenon1T and DarkSide. 
The bound from the neutrino floor restricts the direct detection cross section from below.   
As can be seen in Fig.~\ref{direct}, over the range 0.1 GeV $< m_{\text{DM}} <$ 10 GeV, regions respecting both direct detection bounds and observed relic density are plausible, wherein the respecting values for the scalar mass, $m_\phi$, and the scalar coupling, $\alpha$, are found. Given the viable regions in parameters space which satisfy the 
Xenon1T bound and observed relic density, it can be seen that in these regions there are points with $\alpha$ and $m_\phi$ that can explain the muon anomalous magnetic moment shown as a red band in Fig.~\ref{exclusion}. The future colliders are then capable to 
probe the allowed parameter space we have found here after imposing all the accessible 
bounds.

\section{Conclusion}
We have studied a model with two singlet scalars, one a stable particle as our DM candidate and the other decaying into the SM leptons, to explain the muon anomalous magnetic moment. First we found a secluded region in the parameter space respecting the observed relic abundance. Then we consider the DM-electron elastic scattering cross section being a one-loop process, and constrain the viable space further by applying the direct detection bounds. On the other hand, bounds from the present and future lepton colliders can reach a sensitivity to probe the remaining viable space. 
The scalar mass range $m_\phi \sim 1.5-5$ GeV, is already excluded by {\it BABAR} experiment, 
as can be seen in Fig.~\ref{exclusion}. LDMX (Phase.2) and ILC are capable to reach a sensitivity in order to exclude the rest of singlet scalar mass range respecting the 
muon ($g-2$) anomaly. On the other side, taking the strongest direct detection bounds
there is a large parameter space in the range $m_{\text{DM}} = 0.1-10$ GeV above the 
neutrino floor and the subject of further probe by the future direct detection experiments.

\bibliography{ref}

\bibliographystyle{utphys}

\end{document}